# Oxygen vacancy induced re-entrant spin glass behavior in multiferroic ErMnO$_3$ thin films

S. Y. Jang[1], D. Lee[1], J.-H. Lee[1], T. W. Noh[1], Y. Jo[2], M.-H. Jung[2], and J.-S. Chung[3*]

[1]ReCOE and FPRD, Department of Physics and Astronomy, Seoul National University, Seoul 151-747, Korea

[2]Quantum Material Research Team, Korea Basic Science Institute, Daejeon 305-333, Korea

[3]Department of Physics, Soongsil University, Seoul 156-743, Korea

Epitaxial thin films of hexagonal ErMnO$_3$ fabricated on Pt(111)/Al$_2$O$_3$(0001) and YSZ(111) substrates exhibited both ferroelectric character and magnetic ordering at low temperatures. As the temperature was reduced, the ErMnO$_3$ films first showed antiferromagnetism. At lower temperatures, the films deposited at lower oxygen partial pressures exhibited spin glass behavior. This re-entrant spin glass behavior was attributed to competition between an antiferromagnetic interaction in the hexagonal geometry and a ferromagnetic interaction caused by a change in Mn valence induced by excess electrons from the oxygen vacancies.

Multiferroic materials have attracted considerable attention due to the intriguing interplay between their charge and spin degrees of freedom as well as their potential applications.[1,2] Among the known multiferroic materials, the rare earth manganites $R$MnO$_3$ are intriguing. In bulk forms, $R$MnO$_3$ compounds with smaller $R$ ions (*i.e.*, $R$=Ho-Lu) have hexagonal crystal structures, while those with larger $R$ ions (*i.e.*, $R$=La-Dy) have orthorhombic structures. We recently succeeded in fabricating hexagonal GdMnO$_3$, TbMnO$_3$, and DyMnO$_3$ thin films using epitaxial stabilization techniques.[3-5] These syntheses allowed systematic investigation of the effects of rare earth ion size on the physical properties of the hexagonal $R$MnO$_3$ system.[6]

One of the more intriguing observations is that the epitaxially stabilized hexagonal $R$MnO$_3$ films show antiferromagnetic (AFM) and spin glass (SG) behaviors,[4,5] with HoMnO$_3$ thin films showing similar properties.[7] Note that with the exception of ScMnO$_3$, the SG behaviors are absent in most $R$MnO$_3$ single crystals,[8,9] making it important to understand the origin of these magnetic properties.

Here, we report the successful growth of epitaxial hexagonal ErMnO$_3$ thin films along with a discussion of their magnetic properties. We systematically varied growth conditions, such as the oxygen partial pressure ($P_{O2}$), substrate temperature, laser fluence, *etc.*, and found that the films deposited at lower $P_{O2}$ exhibited SG behaviors. We propose that excess electrons from oxygen vacancies induce ferromagnetic interactions, which result in the SG phenomena by competing with the AFM interaction among the Mn ions.

High quality epitaxial ErMnO$_3$ thin films were grown using a pulsed laser deposition (PLD) technique. The ErMnO$_3$ target was ablated with a 248-nm KrF excimer laser (Lambda Physik) with a laser fluence of 1.5 J/cm$^2$ and a repetition rate of 4 Hz. The substrates were maintained at the optimized temperature of 850°C at $P_{O2}$ ranging from 30-300 mTorr. The deposited film thickness was estimated to be ~100 nm.

The crystal structures were analyzed using both a four-circle high-resolution X-ray diffraction (XRD) machine and those at the 3C2 and 10C1 beamlines at the Pohang Light Source. Figure 1(a) shows the XRD $\theta$-$2\theta$ scans of the ErMnO$_3$ films on the Pt(111)/Al$_2$O$_3$(0001) and (111)-oriented yttria-stabilized zirconia (YSZ) substrates deposited at $P_{O2}$ = 100 mTorr. Only the (0002) and (0004) reflections of the hexagonal ErMnO$_3$ could be seen along with the substrate peaks, indicating the formation of a pure hexagonal phase. The inset of Fig. 1(a) shows that the full width at half maximum in the rocking curve of the ErMnO$_3$ (0002) peak was about 0.176°, indicating a reasonably good crystalline quality film. In figures 1(b) and 1(c), $\phi$-scans of the hexagonal ErMnO$_3$ (10-12) peak indicated a sixfold symmetry, suggesting epitaxial growth of hexagonal ErMnO$_3$ films with epitaxial relationships of ErMnO$_3$[100]||Pt[11-2]||Al$_2$O$_3$[11-20] and ErMnO$_3$[100]||YSZ[11-2], respectively.

Figure 1(d) shows the results of X-ray reciprocal space mapping (RSM) around the YSZ (200) and ErMnO$_3$ (11-22) peaks. The $x$- and $y$-axes correspond to the reciprocal lattice vectors along the in-plane [2-1-1] and out-of-plane [111] directions. If the film is fully strained, the in-plane component of ErMnO$_3$ (11-22) should be matched to that of YSZ (200). From the RSM and $\theta$-$2\theta$ data, the $a$ and $c$ lattice constants were estimated to be 6.14 Å (0.4% tensile strained) and 11.43Å, respectively.

The ferroelectric properties of the ErMnO$_3$ films on Pt(111)/Al$_2$O$_3$(0001) were investigated with a low-temperature probe station (Desert Cryogenics) and a T-F analyzer (aixACCT) at 2 kHz. Similar to other epitaxial hexagonal manganites,[3,4,7] our ErMnO$_3$ thin films showed ferroelectric behavior. Figure 2(a) shows the $P$-$E$ hysteresis loop measured at 140 K. With the maximum applied electric field of 1.8 MV/cm, the remnant polarization and the saturated polarization were about 3.1 μC/cm$^2$ and 11.9 μC/cm$^2$, respectively.

Figure 2(b) shows magnetization *vs.* magnetic field hysteresis loops measured at 5, 40, and 100 K collected with a magnetic property measurement system (Quantum Design). Ferromagnetic hysteresis was observed in the 5 K data, originating from the ferrimagnetic ordering of the Er$^{3+}$ spins.[10] The 100 K data showed a paramagnetic response, indicating that magnetic ordering occurred below 100 K. Raman scattering measurements (not shown here) also indicated that a phonon peak, involving the Mn and O ions in the in-plane, deviates from the dependence expected for anharmonic decay at around 72 K. This temperature corresponded to the antiferromagnetic ordering temperature of Mn ions.[11]

The ErMnO$_3$ thin films deposited at lower P$_{O2}$ exhibited SG behaviors. Figure 3 shows the temperature-dependent *dc* magnetization curves with an applied magnetic field of 100 Oe under zero-field-cooled (ZFC) and field-cooled (FC) conditions. Figures 3(a), (b), (c), and (d) show the *dc* magnetization curves for samples deposited with P$_{O2}$ = 30, 100, 200, and 300 mTorr, respectively. Note that the ZFC and FC curves of the ErMnO$_3$ films deposited at 30 mTorr and 100 mTorr begin to separate from each other below ~45 K. Such *dc* magnetization behavior is an indicator of an SG transition, with a freezing temperature (T$_f$) of ~45 K. The inset in Fig. 3(b) shows the temperature dependence of the initial susceptibility $\chi_0 = (dM/dH)_{H \to 0}$, which generally presents a peak at T$_f$.[12] As shown in the inset, the $\chi_0$ data have a shoulder around 40 K, while the peak at zero temperature is due to ferrimagnetic ordering of the Er$^{3+}$ spins. This shoulder can be regarded as further evidence for the transition into the SG phase. In contrast to the 30- and 100-mTorr data, there were no differences in the ZFC or FC curves for the films deposited at 200 and 300 mTorr.

The SG behavior was most likely due to modulation of the antiferromagnetic ordering of Mn sites in the hexagonal crystal structure, where the spins of Mn ions within the Mn-O plane form a triangular network. To avoid geometric magnetic frustration, the Mn spins are ordered antiferromagnetically with a rotation of 120° below $T_N$.[10] However, disorders or other competing interactions can cause frustration of spins. For example, the excess Mn atoms in Mn-rich hexagonal $YMnO_3$ have been reported to drive the system to a SG ground state below 42 K.[13]

It is interesting that, initially, the Mn spins order antiferromagnetically with $T_N \sim 72$ K for our $ErMnO_3$ films, and then undergo a transition to the SG phase at $T_f \sim 45$ K. Such SG material, which has long-range magnetic ordering above $T_f$, is known as a reentrant SG (RSG) system.[14] Upon cooling, the ordered state should compete with other types of long-range order, such as random ferromagnetic interactions that increase as temperature decreases. As a result, it becomes frustrated and undergoes the RSG transition.

The $P_{O2}$ dependence of the magnetization data suggested that oxygen vacancies inside the thin films may cause the RSG behavior. To confirm the increase in oxygen vacancies, we measured the systematic changes in the lattice parameters for $ErMnO_3$ thin films deposited at different $P_{O2}$ (see Fig. 4(a)). For the films grown at lower $P_{O2}$, the $ErMnO_3$ (0004) peak shifted toward a lower $2\theta$ angle, indicating that the lattice constant *c* increased. But, the lattice constant *a*'s from the RSM data were independent of $P_{O2}$ and almost relaxed, *i.e.*, in 0.38, 0.4, 0.4, and 0.5 % tensile strain for 30, 100, 200, and 300 mTorr samples, respectively, suggesting negligible effect of strain. Therefore, as shown in the inset of Fig. 4(a), the unit cell volume ($\sqrt{3}/2)a^2c$, expanded systematically at lower $P_{O2}$, which could be attributed to the increase in oxygen vacancies within the ErMnO$_3$ layer.

Further insight on the chemical status of Mn ions was obtained from X-ray photoemission spectroscopy (XPS) measurements with an Al anode ($h\nu$ =1486.6 eV). In general, even in the pure Mn valence state, the Mn 2p doublet consists of numerous multiplet peaks,[15] from which it is difficult to extract reasonable information by fitting the peak structure. Therefore, we directly compared the overall features. As shown in Fig. 4(b), the Mn 2p states of the films deposited at lower oxygen partial pressures systematically shifted to lower binding energies as compared to films deposited at higher pressures. This shift was related to a change of valence from Mn$^{3+}$ to Mn$^{2+}$ in the stoichiometric ErMnO$_3$ case.[16] The screening effect caused by an excess d-electron in the Mn$^{2+}$ ion lowers the binding energy of the 2p electrons compared to in the Mn$^{3+}$ ion. XPS revealed that the excess electrons provided by the oxygen vacancies cause more Mn$^{2+}$ states to be formed in the ErMnO$_3$ film deposited at lower $P_{O2}$.

The increase in Mn$^{2+}$ states enhances the double-exchange interaction between Mn$^{2+}$ and Mn$^{3+}$ sites, which favors a ferromagnetic spin configuration.[13,17] The RSG state is formed below $T_N$ when the ferromagnetic interaction becomes strong enough to compete with the antiferromagnetic interaction. It is noted that the earlier reported RSG phase of Mn-rich YMnO$_3$ is induced by excess Mn compared to the situation here where the oxygen vacancies drive the system to the RSG state.

In summary, high-quality epitaxial hexagonal ErMnO$_3$ thin films were fabricated exhibiting both ferroelectric features and magnetic transitions, including antiferromagnetic ordering of Mn$^{3+}$ and ferrimagnetic ordering of Er$^{3+}$. The ErMnO$_3$ films deposited at lower $P_{O2}$ showed a transition to the RSG phase, which was ascribed to competition between the FM and the AFM interaction in the hexagonal structure. In addition, the formation of the RSG state could be controlled by varying $P_{O2}$. The results of the present study will make it possible to obtain a better understanding of SG behavior in epitaxial hexagonal manganite films, which have potential practical importance beyond their scientific interest.

This work was supported financially by the Creative Research Initiatives (Functionally Integrated Oxide Heterostructure) of the Korean Science and Engineering Foundation (KOSEF). J.–S. C. was supported by KRF-2007-314-C00088 and the Soongsil University Research Fund.

**Figure captions**

FIG. 1. (Color online) (a) $\theta$-$2\theta$ XRD scans of ErMnO$_3$ on Pt(111)/Al$_2$O$_3$(0001) and YSZ(111) substrates. The inset in (a) shows a rocking curve measured around the ErMnO$_3$ (004) peak. (b) and (c), $\phi$-scans of the (10-12) peaks for ErMnO$_3$ films on Pt(111)/Al$_2$O$_3$(0001) and YSZ(111) substrates, respectively. The dotted line in (c) is the $\phi$-scan of the YSZ (200) peak. For better comparison, we normalized the intensity of the substrate peaks. (d) Reciprocal space mapping around the (200) Bragg reflection from the YSZ substrate and the (11-22) Bragg reflection from the ErMnO$_3$ films. The map is relative to a YSZ (200) reflection.

FIG. 2. (Color online) (a) Polarization *vs.* electric field hysteresis loop for the ErMnO$_3$ film at 140K. (b) Magnetic field-dependent magnetization curves at 5, 40, and 100 K. The inset shows the Raman peak position of one of the E$_2$ symmetry modes of the ErMnO$_3$ film *vs.* temperature. The solid line shows the peak position expected for a standard anharmonicity-related phonon decay.

FIG. 3. (Color online) Temperature-dependent magnetization curves of ErMnO$_3$ thin films deposited at oxygen partial pressures of (a) 30, (b) 100, (c) 200, and (d) 300 mTorr under ZFC and FC conditions. The inset in (b) shows the temperature dependence of the *initial susceptibility* $\chi_o = (dM/dH)_{H \to 0}$.

FIG. 4. (Color online) (a) $\theta$-$2\theta$ XRD scans of ErMnO$_3$ on YSZ(111) substrates deposited at oxygen partial pressures of 30, 100, 200, and 300 mTorr. The inset shows the calculated unit cell volume ($V_{\text{unit cell}}$) of the films. (b) XPS Mn 2p spectra of the ErMnO$_3$ films deposited at 30 and 300 mTorr. The inset shows the difference between the two spectra.

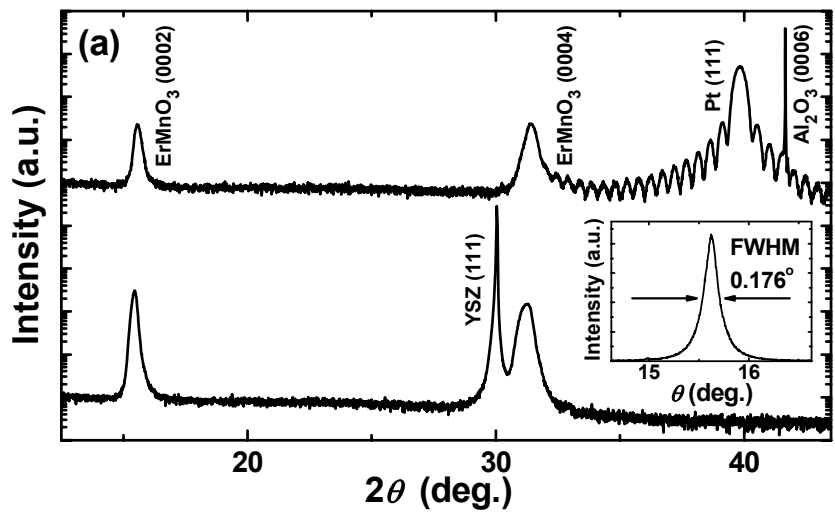
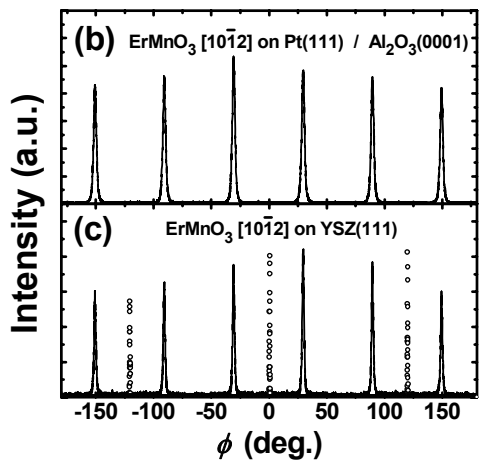
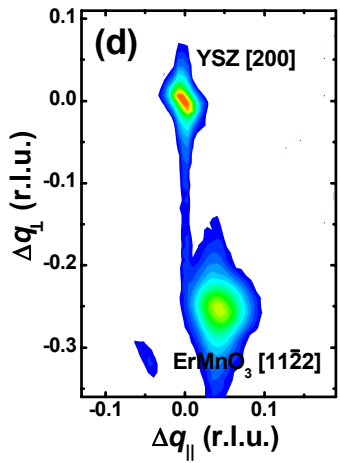

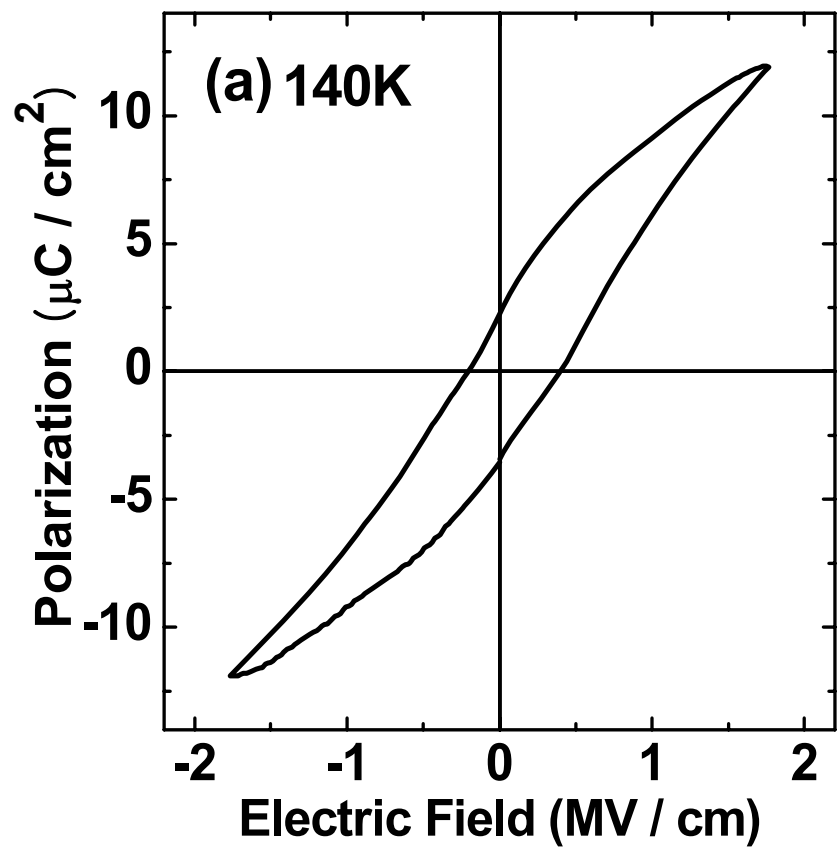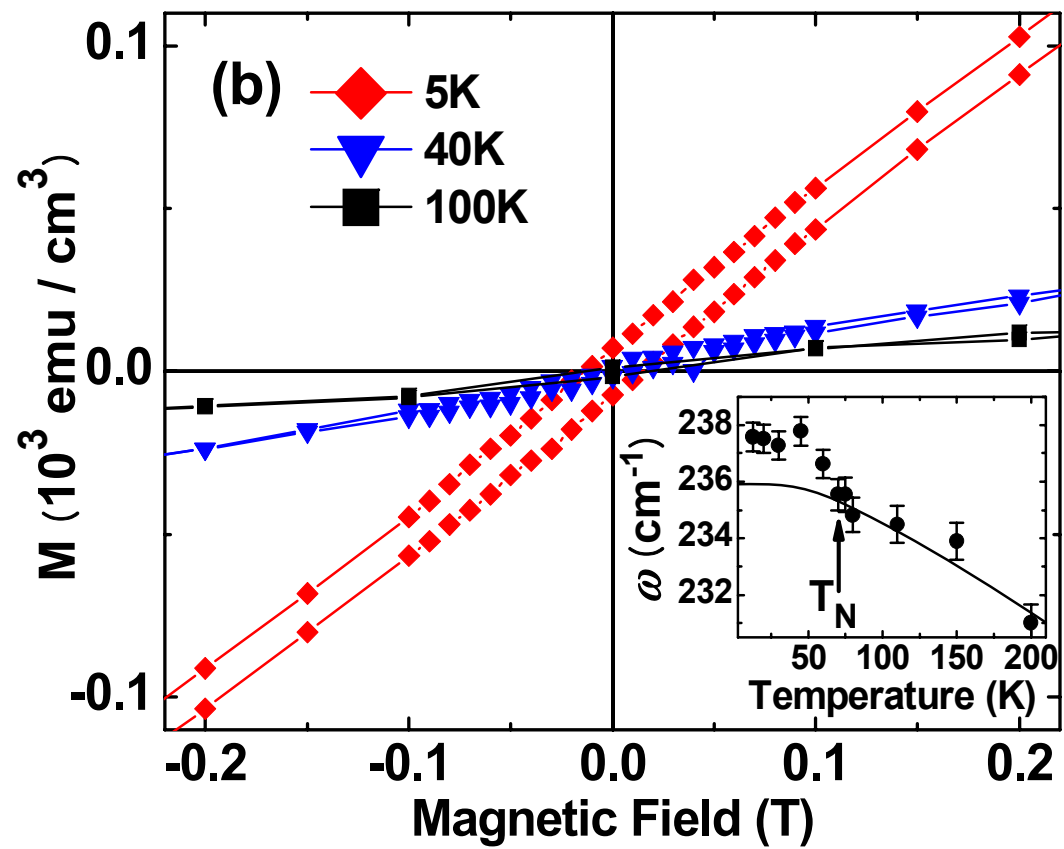

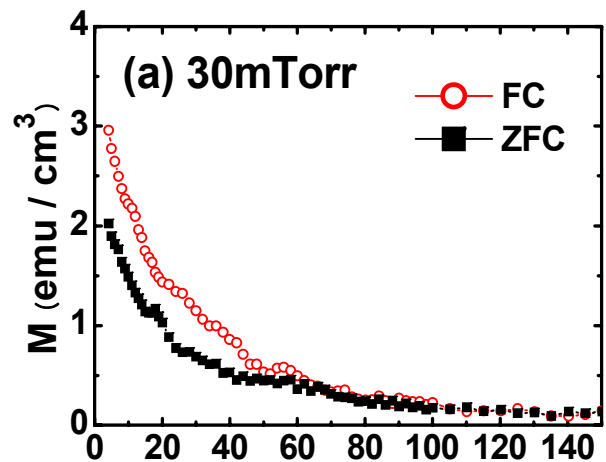

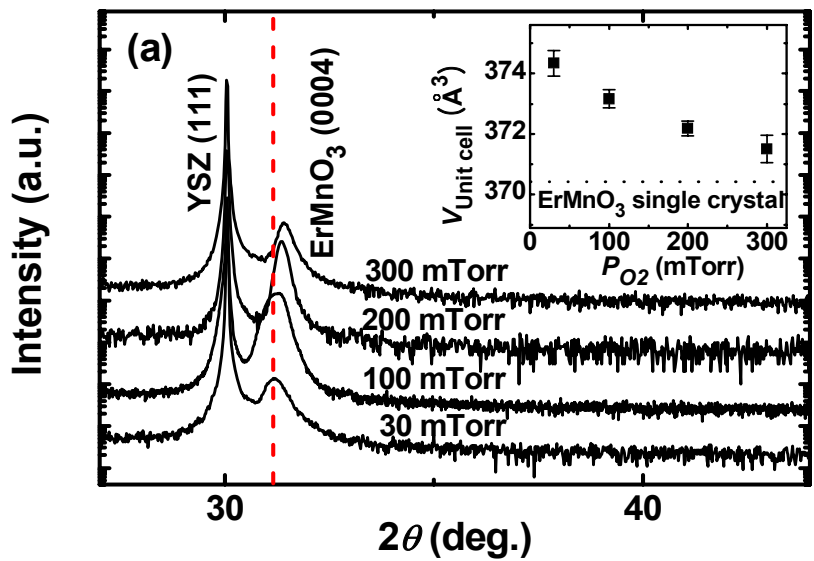
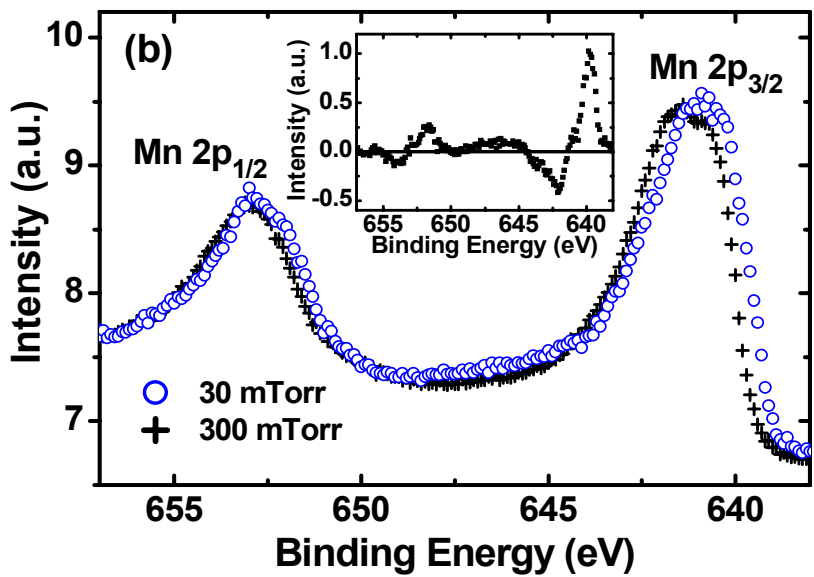